\documentclass[fp,twocolumn]{jpsj3}
\usepackage{amsmath}
\usepackage{txfonts}
\usepackage{braket}
\usepackage{graphicx, color}
\usepackage{here}
\usepackage{nidanfloat}

\begin{document}
\title{ Ground-state and Excitation spectra of Bose-Fermi Mixtures in a Three-Dimensional Optical Lattice}

\author{Rei Hatsuda and Emiko Arahata}
\inst{Faculty of Science and Engineering, Tokyo Metropolitan University,\\
Hachioji, Tokyo $192-0397$, Japan} 
\abst{
Motivated by recent developments in the experimental study of Bose-Fermi mixtures, we investigate ground-state phase diagrams and excitation spectra for Bose-Fermi mixtures in a three-dimensional (3D) optical lattice. 
The Gutzwiller approximation is used to identify a new phase in which both superfluid bosons and metal fermions coexist. 
As a useful probe to identify the quantum phases, we also calculate the excitation spectra. In Mott insulator phase, two excitation features appear in the spectra that correspond to particle and hole excitations.  In superfluid phase, there are Bogoliubov modes and amplitude modes. In coexisting phase, two gapless dispersive modes are identified, which shift due to interaction between bosons and fermions.
}

\maketitle 
\section{INTRODUCTION}
A Bose-Fermi superfluid (SF) mixture in which both Bose and Fermi gases are SFs, was recently realized by the ENS group \cite{siryo-7}. 
This experiment has renewed interest in Bose-Fermi SF mixtures. In particular, quantum phases of Bose-Fermi mixtures in a three-dimensional optical lattice have attracted much attention, from both theoretical and experimental perspectives\cite{siryo-6,siryo-5,siryo-4}. Bose-Fermi mixtures in an optical lattice \cite{siryo-3,siryo-2} can be well described by the so-called tight binding Bose-Fermi-Hubbard model, which was derived in Ref. \cite{siryo-1}.
The model is expected to be useful for new quantum simulators using cold bosonic and fermionic atoms.
Parameters such as the effective interaction between Bose and Fermi gases, the total filling factor, and the number ratio of bosons and fermions can be tuned experimentally \cite{siryo0}. Such variety of the physical parameters can result in many quantum phases. 
In particular, the existence of a coexisting Mott insulator (MI)\cite{siryo1,siryo2} phase has been reported theoretically, in which the total number of bosons and fermions is an integer value but both bosons and fermions take some intermediate fillings\cite{siryo3}. 
However, most theoretical studies have concentrated on the filling factor for bosons and fermions in optical lattices, so that the effects of tuning the other parameters have not yet been determined in any detail. 
In Ref.[11], the phase diagrams of Bose-Fermi mixtures in a 3D optical lattice were investigated by dynamical mean field theory (DMFT) with limited interaction parameters. 
However, it is necessary to consider tuning of the variational parameters to produce novel phase diagrams.
To understand such novel phases, it is important to reveal the details of Bose-Fermi mixture excitations in the 3D optical lattice.
The excitations of the Bose-Fermi-Hubbard model from various quantum phases, including SF and MI, have been well studied theoretically\cite{siryo4,siryo5,siryo6}. The excitation spectra have been observed experimentally via Bragg spectroscopy. The experimental results were in good agreement with the theoretical results and the phases have been successfully classified. It is thus expected that elementary excitations of Bose-Fermi mixtures will be observed and that different phases could be characterized.
However, a simpler method is required to investigate phase diagrams and excitations with the tuning of variational parameters. \\
\ \ \ In this work, we study the quantum phases and excitation properties of a Bose-Fermi mixture in a 3D optical lattice at zero temperature using the Gutzwiller approximation\cite{siryo7,siryo8,siryo9} by the tuning of variational parameters. We first determine the ground-state phase diagrams to identify phases that should be focused on. Second, we investigate excitations by extending the method for the calculation of excitation spectra of Bose-Fermi mixtures. The Gutzwiller approximation is effective for performing ground state studies of many important phenomena. 
This approximation is known to be valid when the spatial dimensions are high enough, Bose-Bose interactions are not weak and the Bose-Fermi interactions are not very strong.
For these cases, Lin et al. applied the Gutzwiller approximation to the interacting Bose-Fermi-Hubbard model and compared the results with that by DMFT \cite{siryo3}, and both were found to be in good agreement. \\ 
\ \ \ 
The excitations of a single-component Bose-Hubbard model are well understood from previous theoretical and experimental studies\cite{siryo10}. The lowest two branches of the excitation spectrum in a MI phase correspond to the particle- and hole-excitation modes. 
In a SF phase, one gapless mode and gap-containing mode appear. The gapless mode is known as  Bogoliubov mode, while the lowest gap-containing mode is an amplitude mode. \\ 
\ \ \ This paper is organized as follows. In Sec. II, we explain formulations based on
the Bose-Fermi-Hubbard model, the Gutzwiller approximation and the linearized equations of motion. 
 In Sec. III, we show the phase diagrams for several parameters using the formulation described in Sec. II. We find a new phase including coexistence of SF bosons and metal fermions.
In Sec. IV, we calculate excitation spectra for each of the phases obtained in Sec. III.
We show several changes in the excitation spectra due to the interactions between the bosons and the fermions. Conclusions are given in Sec V. 
\section{FORMULATION}
We consider Bose-Fermi mixtures in a 3D  optical lattice which is well described by the Bose-Fermi-Hubbard model. 
\subsection{Gutzwiller approximation}
\ \ Here, fermions are supposed to be spinless. Ignoring an external trapping potential, the Bose-Fermi-Hubbard Hamiltonian is given by: 
\begin{align}
\hat{H}_{\rm b}&=-t_{\rm b}\sum_{<i,j>}b_{i}^{\dagger}{b}_{j}+\frac{U_{\rm bb}}{2}\sum_{i}n_{{\rm b}i}(n_{{\rm b}i}-1)-\mu_{\rm b}\sum_{i}n_{{\rm b}i},\\
\hat{H}_{\rm f}&=-t_{\rm f}\sum_{<i,j>}c_{i}^{\dagger}{c}_{j}-\mu_{\rm f}\sum_{i}n_{{\rm f}i},\\
\hat{H}_{\rm bf}&={U_{\rm bf}}\sum_{i}n_{{\rm b}i}n_{fi}.
\end{align}
The subscripts b and f denote bosons and fermions, respectively. 
$b_{i}^{\dagger}$ ($c_{i}^{\dagger}$) and $b_{i}$ ($c_{i}$) represent bosonic (fermionic) creation and annihilation operators at site i, respectively. 
$<i,j>$ in the first term of $\hat{H}_{\rm b}$ and $\hat{H}_{\rm f}$ denote the sum over nearest neighbors.
$n_{{\rm b}i}$ and $n_{{\rm f}i}$ are the respective occupation numbers of bosons and fermions at site i. 
${U_{\rm bb}}$ and ${U_ {\rm bf}}$ are boson-boson and boson-fermion on-site interactions, respectively. $\mu_{\rm b}$ and $\mu_{\rm f}$ are the chemical potentials of bosons and fermions, respectively. $t_{\rm b}$ and $t_{\rm f}$ are the hopping energies of bosons and fermions, respectively. $\hat{H}_{\rm b}$ and $\hat{H}_{\rm f}$ are purely bosonic and fermionic Hamiltonians, respectively.
$\hat{H}_{\rm bf}$ describes the boson-fermion on-site interactions.
\ \ The Gutzwiller approximation is used to investigate the ground state of the Bose-Fermi mixtures. 
The Gutzwiller-type variational wave function is assumed to take a simple form of:  \\
\begin{align}
\Ket{\psi_{G}}=\prod_{i}\sum_{n_{\rm b},n_{\rm f}}f^{(i)}_{n_{\rm b},n_{\rm f}}\Ket{n_{\rm b},n_{\rm f}}_{i},
\end{align}
where $\Ket{n_{\rm b},n_{\rm f}}_{i}$ is the Fock state with the average number of bosons $n_{\rm b}$ and fermions $n_{\rm f}$. The variational factors $f^{(i)}_{n_{\rm b},n_{\rm f}}$ satisfy the normalization condition, 
$
\sum_{n_{\rm b},n_{\rm f}}\|f^{(i)}_{n_{\rm b},n_{\rm f}}\|^{2}=1
$. 
The expectation value of the Bose-Fermi Hubbard Hamiltonian can be evaluated as:
\begin{align}
&E_{n_{\rm b},n_{\rm f}}=\Braket{\psi_{G}|H|\psi_{G}} \notag \\
&=\sum_{i}\sum_{n_{\rm b},n_{\rm f}}\left(\frac{U_{\rm bb}}{2}n_{\rm b}\left(n_{\rm b}-1\right)+U_{\rm bf}n_{\rm b}n_{\rm f}-\mu_{\rm b}n_{\rm b}-\mu_{\rm f}n_{\rm f}+t_{f}Z_{f}\epsilon \right) \notag \\ 
&\times\|f^{(i)}_{n_{\rm b},n_{\rm f}}\|^{2} -t_{b}\sum_{<i,j>}\left(\phi_{i}^{\ast}\phi_{j}+\phi_{j}^{\ast}\phi_{i}\right)
\end{align}
where SF parameter $\phi_{i}$ is given by
\begin{align} 
\phi_{i}=\sum_{n_{\rm b},n_{\rm f}}f^{\ast(i)}_{n_{\rm b},n_{\rm f}}f^{(i)}_{n_{\rm b}+1,n_{\rm f}}. 
\end{align}
Here we can approximate 
$
\Braket{c_{i}^{\dagger}c_{j}} \approx Z_{\rm f}\Braket{c_{i}^{\dagger}c_{j}}_{0} 	\equiv Z_{\rm f}\epsilon. 
$ 
, where
\begin{align}
&Z_{\rm f} 
=\sqrt{\frac{1}{n_{\rm fi}(1-n_{\rm fi})}}\sum_{n_{\rm b},n_{\rm f}}f^{\ast(i)}_{n_{\rm b},n_{\rm f}}f^{(i)}_{n_{\rm b},n_{\rm f}+1}\sqrt{\frac{1}{n_{\rm fj}(1-n_{\rm fj})}}\sum_{n_{\rm b},n_{\rm f}}f^{\ast(j)}_{n_{\rm b},n_{\rm f}}f^{(j)}_{n_{\rm b},n_{\rm f}+1}.
\end{align}
\ \ \ The quasi-particle weight $Z_{\rm f}$ denotes the strength of the correlation between fermions, and $\epsilon$ is the kinetic energy of the non-interacting fermions. \\

By minimizing the effective action $\int d\tau\Braket{\psi_{G}|i\hbar\frac{d}{dt}-H|\psi_{G}}$ with respect to $f^{\ast(i)}_{n_{\rm b},n_{\rm f}}$ , $f'{}s$ equation is obtained as:
\begin{align}
i\hbar\frac{df^{(i)}_{n_{\rm b},n_{\rm f}}}{dt}=\frac{\partial E_{n_{\rm b},n_{\rm f}}}{\partial f^{\ast(i)}_{n_{\rm b},n_{\rm f}}}
\end{align}
and it leads to the Gutzwiller equation: 
\begin{align}
&i\hbar\frac{df^{(i)}_{n_{\rm b},n_{\rm f}}}{dt} \notag \\
&=\left(\frac{U_{\rm bb}}{2}n_{\rm b}
\left(n_{\rm b}-1\right)+U_{\rm bf}n_{\rm b}n_{\rm f}-\mu_{\rm b}n_{\rm b}-\mu_{\rm f}n_{\rm f}
\right)f^{(i)}_{n_{\rm b},n_{\rm f}} \notag\\
&-t_{\rm b}\sum_{j}\phi_{j}\sqrt{n_{\rm b}}f^{(i)}_{n_{\rm b}-1,n_{\rm f}}-t_{\rm b}\sum_{j}\phi_{j}^{\ast}\sqrt{n_{\rm b}+1}f^{(i)}_{n_{\rm b}+1,n_{\rm f}}\notag \\
&+t_{\rm f}\sum_{j}\sqrt{Z_{\rm fj}}\epsilon f^{(i)}_{n_{\rm b},n_{\rm f}-1}+t_{\rm f}\sum_{j}\sqrt{Z_{\rm fj}^{\ast}}\epsilon f^{(i)}_{n_{\rm b},n_{\rm f}+1}.
\end{align}
We can obtain the ground state by the imaginary time propagation method using the imaginary time $t=i\tau$.

\subsection{Linearized equations of motion}
\ \ We now consider the energy and excitation spectra for each quantum phase obtained using the Gutzwiller equation.
We assume a small fluctuation around the stationary variational parameter $\tilde{f}$: 
\begin{eqnarray}
f^{(i)}_{n_{\rm b},n_{\rm f}} =\left(\tilde{f}^{(i)}_{n_{\rm b},n_{\rm f}}+\delta f^{(i)}_{n_{\rm b},n_{\rm f}}\right)\mathrm{e}^{-i\tilde{\omega}_{i}t}.
\end{eqnarray}
We expand the small fluctuation in terms of a plane wave: 
\begin{eqnarray}
\delta f^{(i)}_{n_{\rm b},n_{\rm f}}
=\sum_{\mathbf{k}}\left(u^{(i)}_{n_{\rm b},n_{\rm f},\mathbf{k}}\mathrm{e}^{i\left(\mathbf{k}\cdot\mathbf{r_{i}}-\omega_{k}t\right)}
-v^{\ast(i)}_{n_{\rm b},n_{\rm f},\mathbf{k}}\mathrm{e}^{-i\left(\mathbf{k}\cdot\mathbf{r_{i}}-\omega_{k}t\right)}\right),
\end{eqnarray}
where $\mathbf{r_{i}}$ is the position vector of site i . We then obtain 
\begin{eqnarray*}
\begin{split}
\omega_{\mathbf{k}}u_{n_{\rm b},n_{\rm f},\mathbf{k}} 
&=\left(\frac{U_{\rm bb}}{2}n_{\rm b}\left(n_{\rm b}-1\right)+U_{\rm bf}n_{\rm b}n_{\rm f} \right.\\
&\left.-\mu_{\rm b}n_{\rm b}-\mu_{\rm f}n_{\rm f}+Z_{f}\epsilon+\phi^{2}-\tilde{\omega_{i}}\right)u_{n_{\rm b},n_{\rm f},\mathbf{k}} 
\end{split}
\end{eqnarray*}
\begin{eqnarray}
-\tilde{\phi}\sqrt{n_{\rm b}}u_{n_{\rm b}-1,n_{\rm f},\mathbf{k}}-\tilde{\phi}^{\ast}\sqrt{n_{\rm b}+1}u_{n_{\rm b}+1,n_{\rm f},\mathbf{k}} \nonumber
\end{eqnarray}
\begin{eqnarray}
-\epsilon\left(\mathbf{k}\right)&\left[\displaystyle\sum_{m}\sqrt{m}\sqrt{n_{\rm b}+1}\tilde{f}^{\ast}_{m,n_{\rm f}}\tilde{f}_{n_{\rm b}+1,n_{\rm f}}u_{m-1,n_{\rm f},\mathbf{k}} \right. \nonumber \\
&\left.+\displaystyle\sum_{m}\sqrt{m}\sqrt{n_{\rm b}}\tilde{f}^{\ast}_{m-1,n_{\rm f}}\tilde{f}_{n_{\rm b}-1,n_{\rm f}}u_{m,n_{\rm f},\mathbf{k}} \right. \nonumber  \\
&\left. -\displaystyle\sum_{m}\sqrt{m}\sqrt{n_{\rm b}+1}\tilde{f}_{m-1,n_{\rm f}}\tilde{f}_{n_{\rm b}+1,n_{\rm f}}v_{m,n_{\rm f},\mathbf{k}} \right.\nonumber \\
&\left.-\displaystyle\sum_{m}\sqrt{m}\sqrt{n_{\rm b}}\tilde{f}_{m,n_{\rm f}}\tilde{f}_{n_{\rm b}-1,n_{\rm f}}v_{m-1,n_{\rm f},\mathbf{k}}\right], 
\end{eqnarray}
\begin{eqnarray*}
\begin{split}
\omega_{\mathbf{k}}v_{n_{\rm b},n_{\rm f},\mathbf{k}}&=-\left(\frac{U_{\rm bb}}{2}n_{\rm b}\left(n_{\rm b}-1\right)+U_{\rm bf}n_{\rm b}n_{\rm f} \right. \\
&\left.-\mu_{\rm b}n_{\rm b}-\mu_{\rm f}n_{\rm f}+Z_{f}\epsilon+\phi^{2}-\tilde{\omega_{i}}\right)v_{n_{\rm b},n_{\rm f},\mathbf{k}} 
\end{split}
\end{eqnarray*}
\begin{eqnarray}
+\tilde{\phi}^{\ast}\sqrt{n_{\rm b}}v_{n_{\rm b}-1,n_{\rm f},\mathbf{k}}+\tilde{\phi}\sqrt{n_{\rm b}+1}v_{n_{\rm b}+1,n_{\rm f},\mathbf{k}} \nonumber
\end{eqnarray}
\begin{eqnarray}
+\epsilon\left(\mathbf{k}\right)&\left[\displaystyle\sum_{m}\sqrt{m}\sqrt{n_{\rm b}+1}\tilde{f}_{m,n_{\rm f}}\tilde{f}^{\ast}_{n_{\rm b}+1,n_{\rm f}}u_{m-1,n_{\rm f},\mathbf{k}} \right. \nonumber \\
&\left.+\displaystyle\sum_{m}\sqrt{m}\sqrt{n_{\rm b}}\tilde{f}_{m-1,n_{\rm f}}\tilde{f}^{\ast}_{n_{\rm b}-1,n_{\rm f}}u_{m,n_{\rm f},\mathbf{k}} \right. \nonumber \\
&\left. -\displaystyle\sum_{m}\sqrt{m}\sqrt{n_{\rm b}+1}\tilde{f}^{\ast}_{m-1,n_{\rm f}}\tilde{f}^{\ast}_{n_{\rm b}+1,n_{\rm f}}u_{m,n_{\rm f},\mathbf{k}}
\right. \nonumber \\
&\left.-\displaystyle\sum_{m}\sqrt{m}\sqrt{n_{\rm b}}\tilde{f}^{\ast}_{m,n_{\rm f}}\tilde{f}^{\ast}_{n_{\rm b}-1,n_{\rm f}}u_{m-1,n_{\rm f},\mathbf{k}}\right],
\end{eqnarray}
where $\tilde{\omega}$ is the energy of the stationary state and we define $\epsilon\left(\mathbf{k}\right)=\cos \left(\mathbf{k}\right)$. 
Throughout this paper, we assume $n_{\rm bmax}=4$ and $n_{\rm fmax}=1$.

\section{PHASE DIAGRAMS}
\begin{figure}[H]
\begin{center}
\begin{tabular}{cc}
\begin{minipage}{0.9\hsize}
\centering
\includegraphics[width=1.0\hsize]{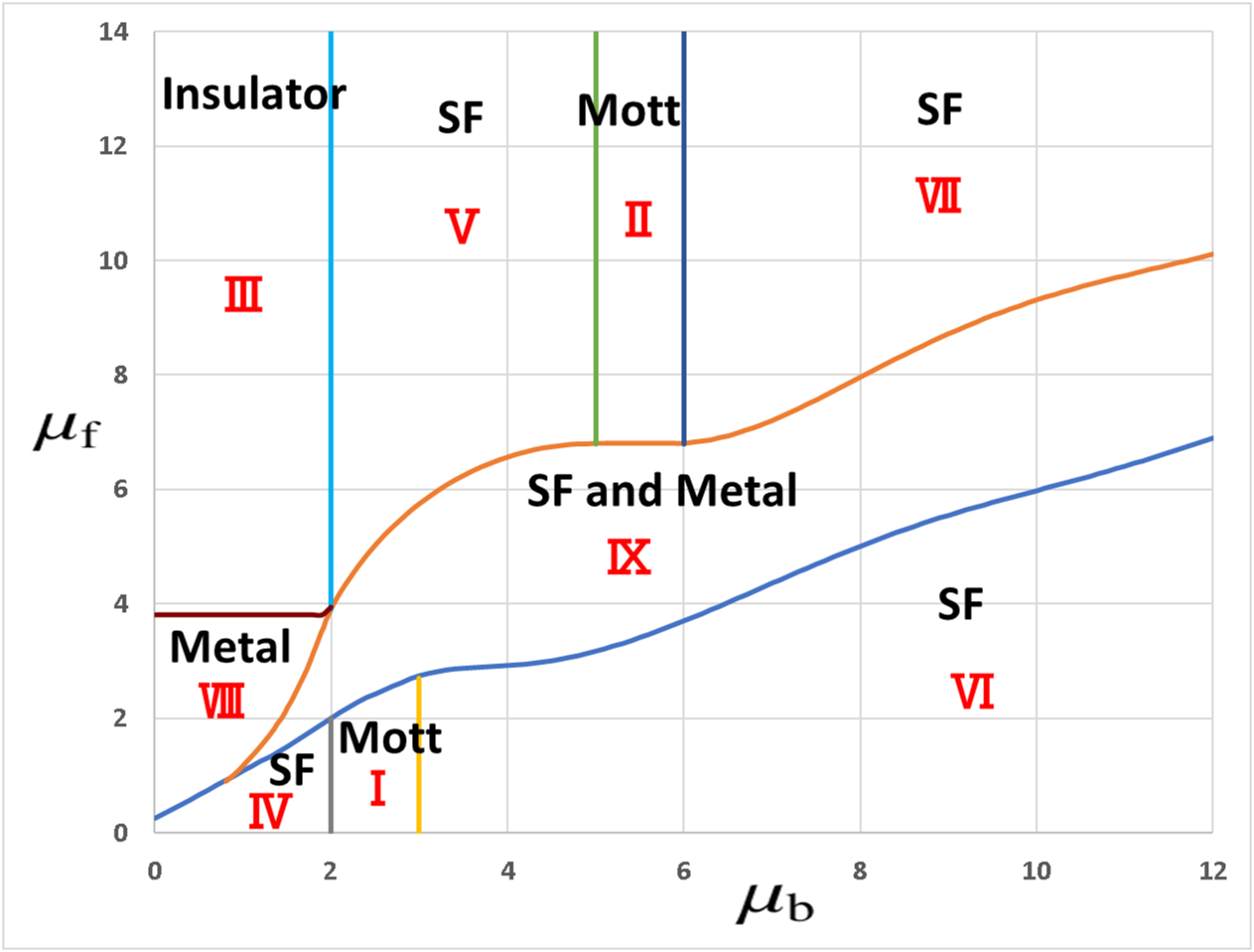}
\hspace{1.6cm} (a)
\end{minipage}\\
\begin{minipage}{0.9\hsize}
\centering
\includegraphics[width=1.0\hsize]{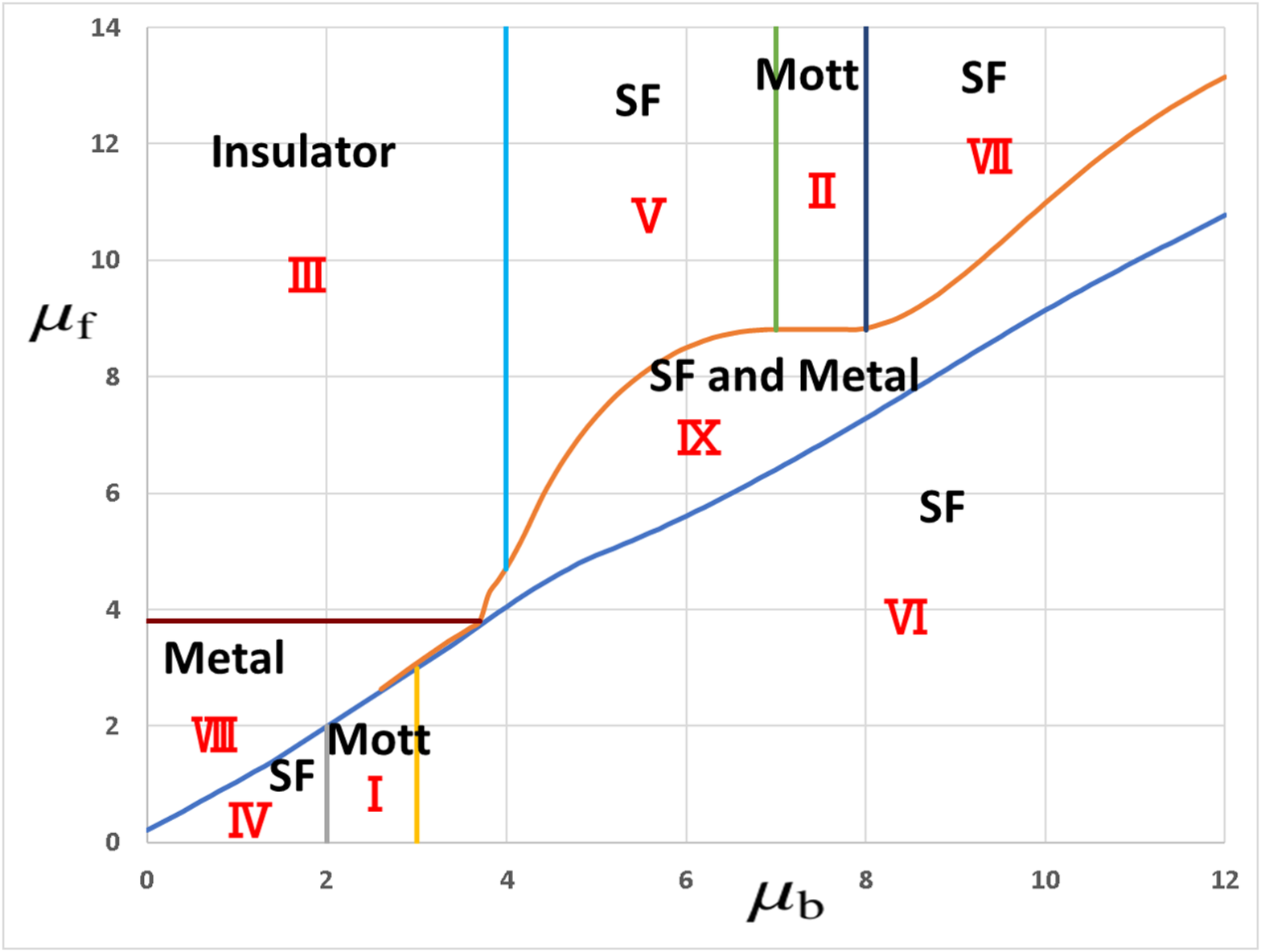}
\hspace{1.6cm} (b)
\end{minipage}
\end{tabular}
\end{center}
\end{figure}

\begin{figure}[H]
\begin{center}
\begin{tabular}{cc}
\begin{minipage}{0.9\hsize}
\centering
\includegraphics[width=1.0\hsize]{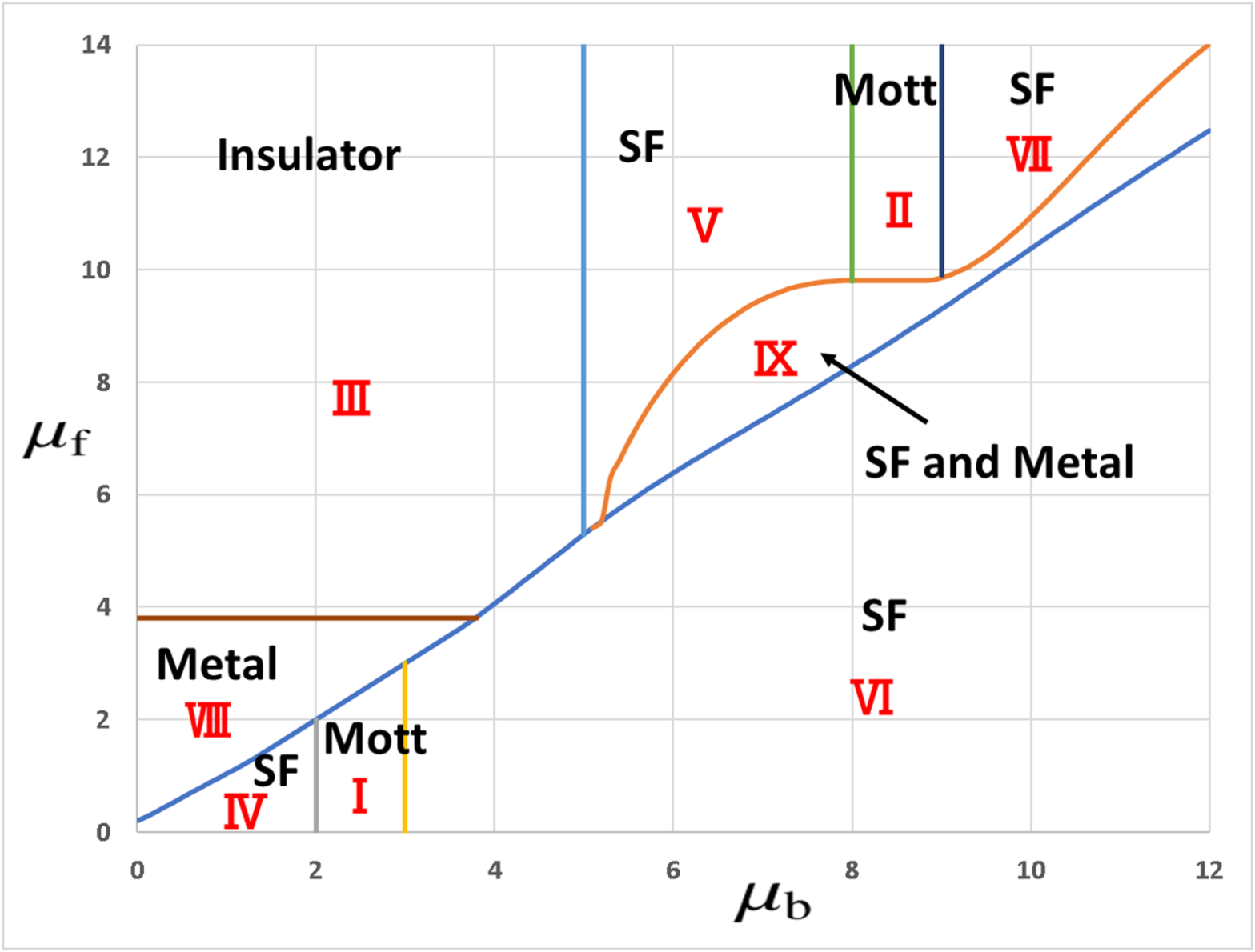}
\hspace{1.6cm} (c)
\end{minipage}\\
\begin{minipage}{0.9\hsize}
\centering
\includegraphics[width=1.0\hsize]{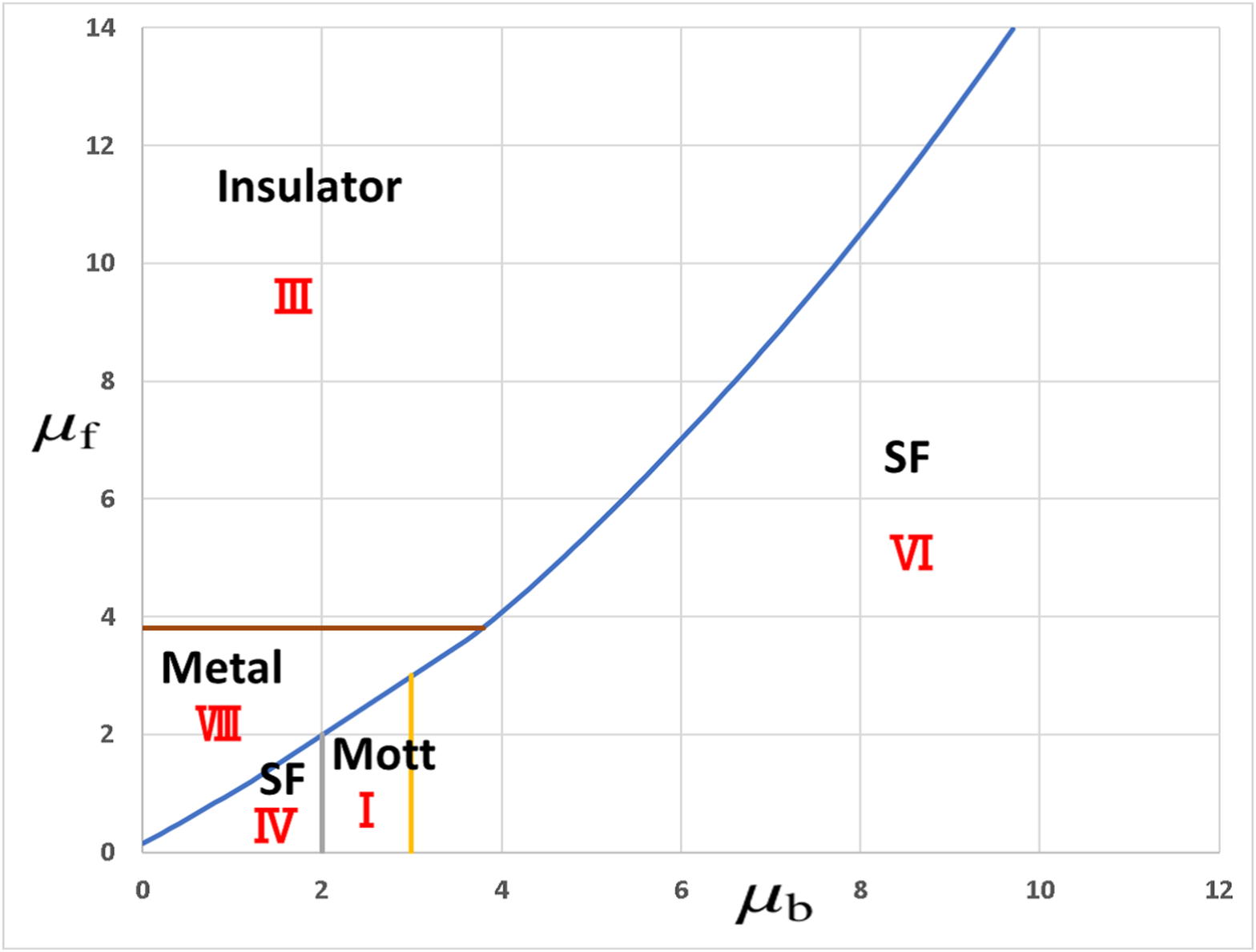}
\hspace{1.6cm} (d)
\end{minipage}
\end{tabular}
\caption{(Color online) Phase 
diagrams in Bose-Fermi mixture spanned by the chemical potentials $\mu_{\rm b}$ and $\mu_{\rm f}$ 
for (a) $U_{\rm bb}=6$, $U_{\rm bf}=3$, (b) $U_{\rm bb}=6$, $U_{\rm bf}=5$, (c) $U_{\rm bb}=6$, $U_{\rm bf}=6$, and (d) $U_{\rm bb}=6$, $U_{\rm bf}=12$.
Region I represents the MI phase of pure bosons. Region II represents the MI phase of bosons and fermions.
Region III represents the insulator phase of pure fermions. Region IV represents the SF phase of pure bosons ($n_{\rm b}<1$).
Region V represents the SF phase of bosons ($n_{\rm b}<1$) and insulator fermions. Region VI represents the SF phase of pure bosons ($n_{\rm b}>1$).
Region VII represents the SF phase of bosons ($n_{\rm b}>1$) and insulator fermions. Region VIII represents the metal phase of pure fermions.
Region IX represents the coexisting phase of SF bosons and metal fermions.
}
\end{center}
\end{figure}

\ In this section we discuss the ground state phase diagrams for the Bose-Fermi mixture in a 3D optical lattice.
We apply a mean field approximation to the bosonic field $\phi_{i}$ and the weight $Z_{\rm f}$ and we assume $zt_{\rm b}=zt_{\rm f}=1$, where $z$ is the coordination number of particle. 
Figure 1 shows the phase diagrams for the Bose-Fermi mixture. 
The interactions are chosen as (a)\ \ $U_{\rm bb}=6$ and $U_{\rm bf}=3$,\ \ (b)\ \ $U_{\rm bb}=6$ and $U_{\rm bf}=5$,\ \ 
(c)\ \ $U_{\rm bb}=6$ and $U_{\rm bf}=6$, and \ \ (d)\ \ $U_{\rm bb}=6$ and $U_{\rm bf}=12$.
The phase diagrams are spanned by the chemical potentials $\mu_{\rm b}$ and $\mu_{\rm f}$.
Nine regions labeled by I-IX represent (I) MI phase with $n_{\rm b}=1$, (II) MI phase with $n_{\rm b}=1$
 and $n_{\rm f}=1$,
(III) insulator phase with $n_{\rm f}=1$, (IV) SF phase with fractional $n_{\rm b}$ and $n_{\rm f}<1$,
(V) SF phase with $n_{\rm b}<1$ and $n_{\rm f}=1 $, (VI) SF phase with $n_{\rm b}>1$,
(VII) SF phase with $n_{\rm b}>1$ and $n_{\rm f}=1$, (VIII) metal phase with fractional $n_{\rm f}<1$ and 
(IX) coexisting phase of SF bosons and metal fermions. As $\mu_{\rm f}$ and $\mu_{\rm b}$ increase, $n_{\rm f}$ and $n_{\rm b}$ increase in (I)-(VIII).\\
\ \ The coexisting phase shrinks as the Bose-Fermi interaction $U_{\rm bf}$ increases. 
Finally, the coexisting phase disappears in Fig. 1(d) at $U_{\rm bb}=6$ and $U_{\rm bf}=12$.
This indicates that the interactions are too strong for the coexistence of SF bosons and metal fermions to persist.
Region III expands with $U_{\rm bf}$ because as the interaction increases, the energy required to add bosons increases.

\section{EXCITATION SPECTRA}

In Sec. II, we showed the formulation for calculating the elementary excitations for a Bose-Fermi mixture in a 3D optical lattice.
The elementary excitation spectra are calculated in this section on the basis of the ground state phase diagrams obtained in Sec. III. Throughout the paper, we assume the momentum of excitations to be $k_x = k_y = k_z=k$.
\subsection{MI phase}

Figure 2 shows the Bose excitation spectra for the MI phase 
in region I for $\mu_{\rm b}=2.8$ and region II for $\mu_{\rm b}=8.2$ in Fig. 1(c).
Since region I contains no fermions and region II contains insulator fermions, the excitation of the fermions can be ignored.
For the bosons, two gap-containing dispersive modes are observed.
The blue line corresponds to Bose particle excitation, which adds a Bose particle to the MI phase,  and the red line corresponds to Bose hole excitation, which removes a Bose particle from the MI phase.
The two dispersive modes are consistent with those of a pure Bose gas.
\begin{figure}[H]
\begin{center}
\begin{tabular}{c}
\begin{minipage}{0.7\hsize}
\centering
\includegraphics[width=1.0\hsize]{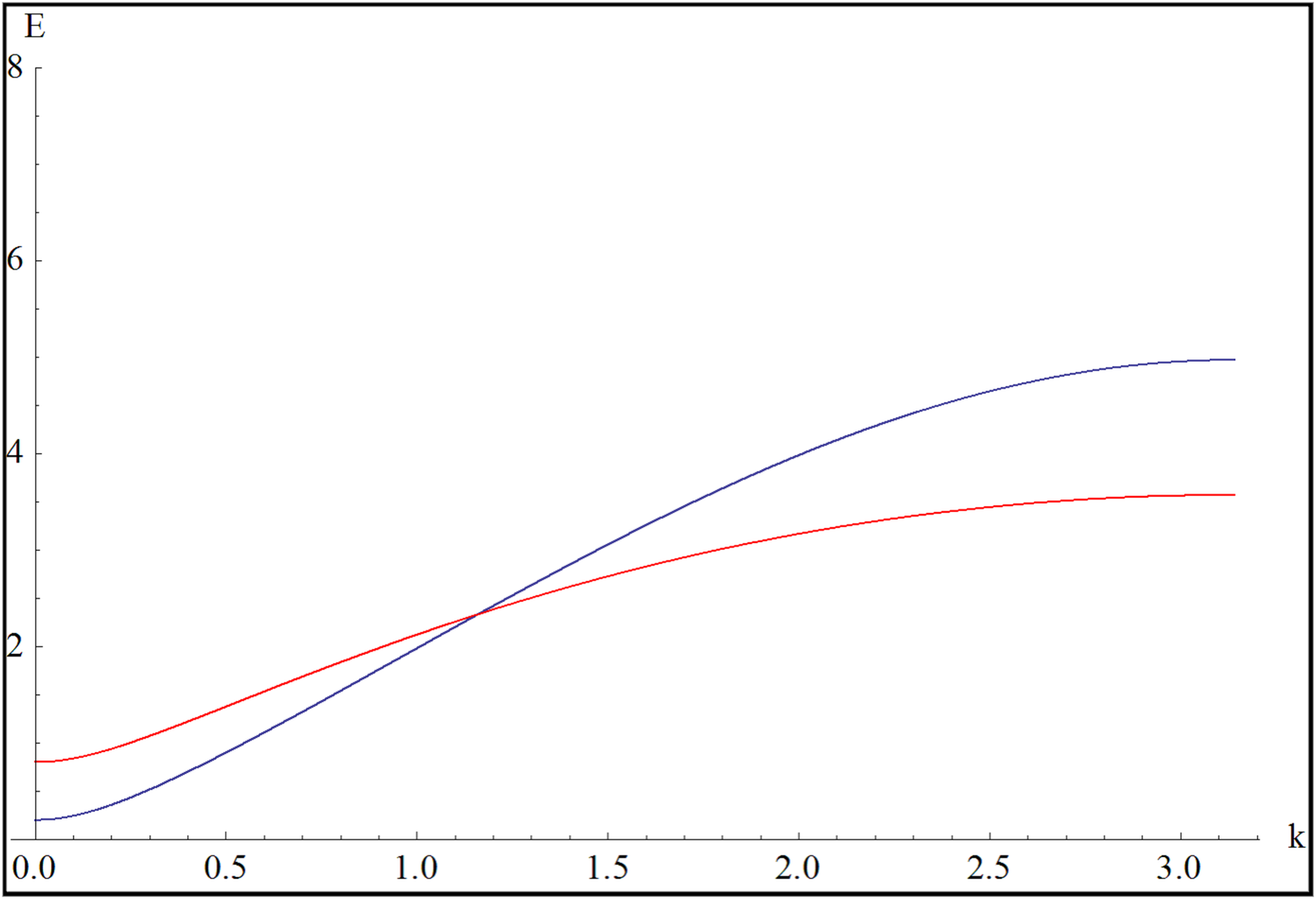}
\hspace{1.6cm} (a)
\end{minipage}\\
\begin{minipage}{0.7\hsize}
\centering
\includegraphics[width=1.0\hsize]{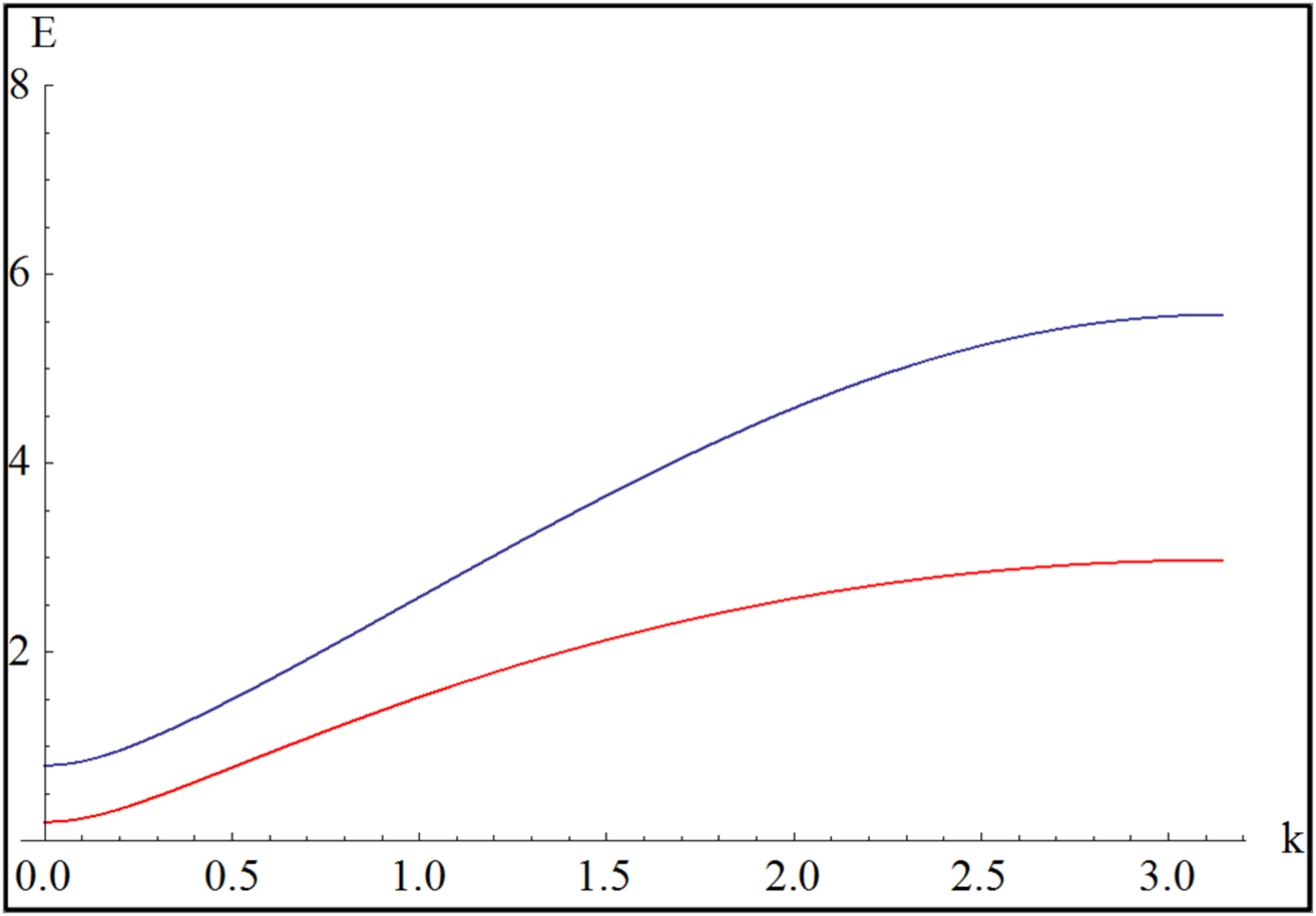}
\hspace{1.6cm} (b)
\end{minipage}
\end{tabular}
\caption{(Color online) Bose excitation spectra of the MI phase 
in (a) region I for $\mu_{\rm b}=2.8$ and (b) region II for $\mu_{\rm b}=8.2$ in Fig. 1(c).
The blue line corresponds to Bose particle excitation and the red line corresponds to Bose hole excitation.
 }
\end{center}
\end{figure}

\subsection{SF phase}
We now consider the SF phase. 
Figure 3 shows excitation spectra of the SF phase of the bosons in (a) region IV for $\mu_{\rm b}=1$, $\mu_{\rm f}=0.5$, (b) region V for $\mu_{\rm b}=6$, $\mu_{\rm f}=14$, (c) region VI for $\mu_{\rm b}=10$, $\mu_{\rm f}=6$, and (d) region VII for $\mu_{\rm b}=10$, $\mu_{\rm f}=16$.

\begin{figure}[H]
\begin{center}
\begin{tabular}{cc}
\begin{minipage}{0.9\hsize}
\centering
\includegraphics[width=\hsize]{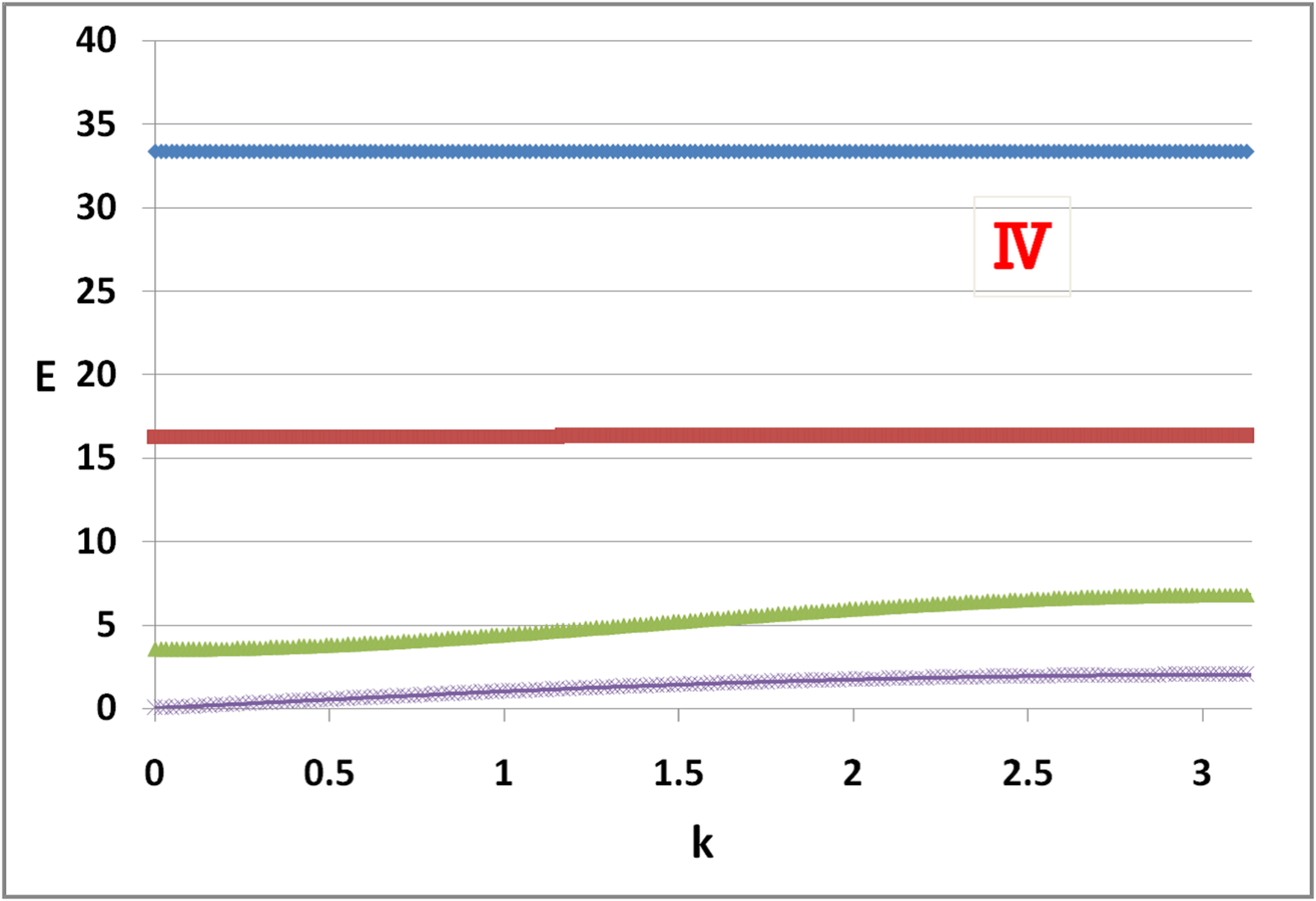}
\hspace{1.6cm} (a)
\end{minipage}\\
\begin{minipage}{0.9\hsize}
\centering
\includegraphics[width=\hsize]{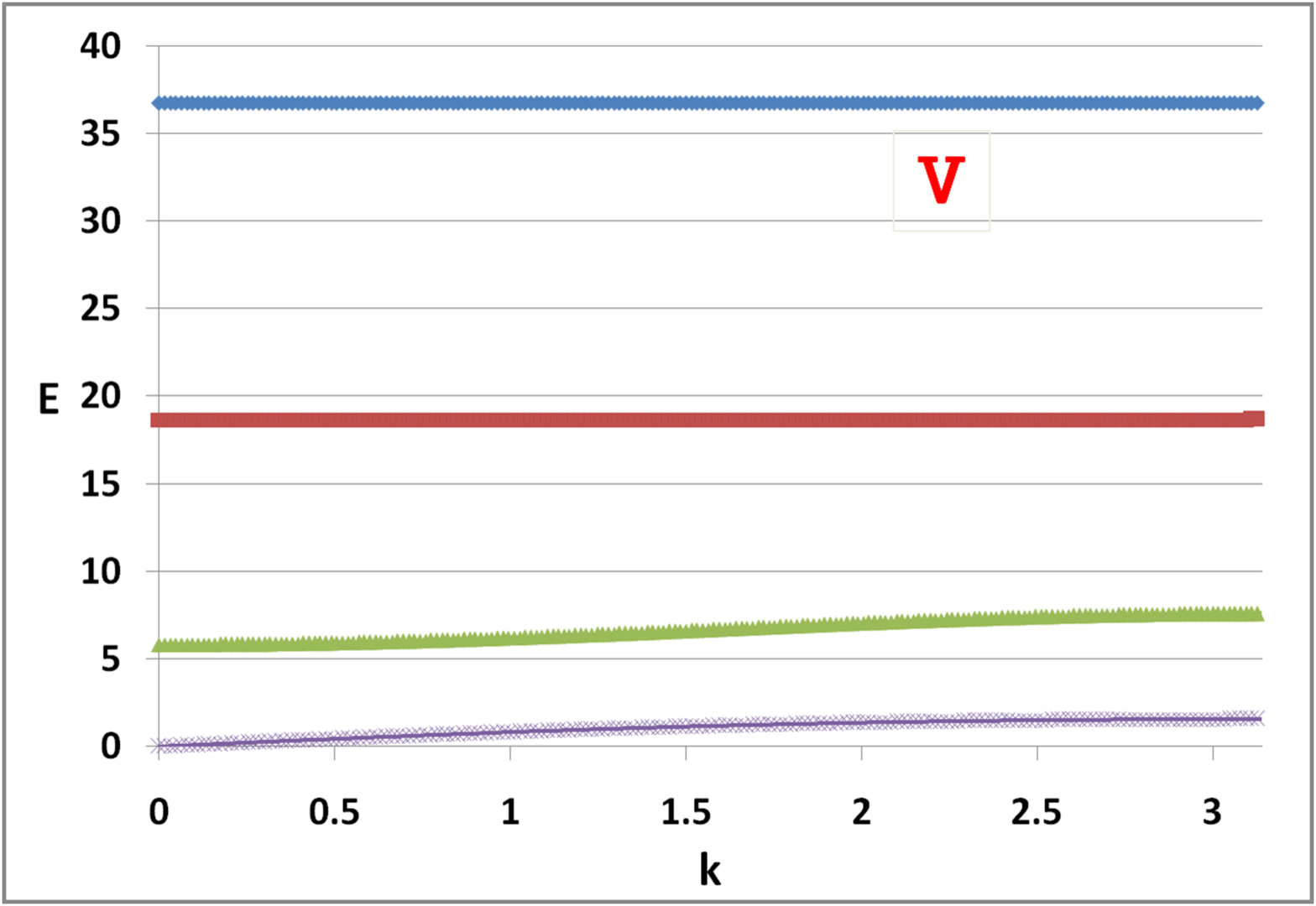}
\hspace{1.6cm} (b)
\end{minipage}\\
\begin{minipage}{0.9\hsize}
\centering
\includegraphics[width=\hsize]{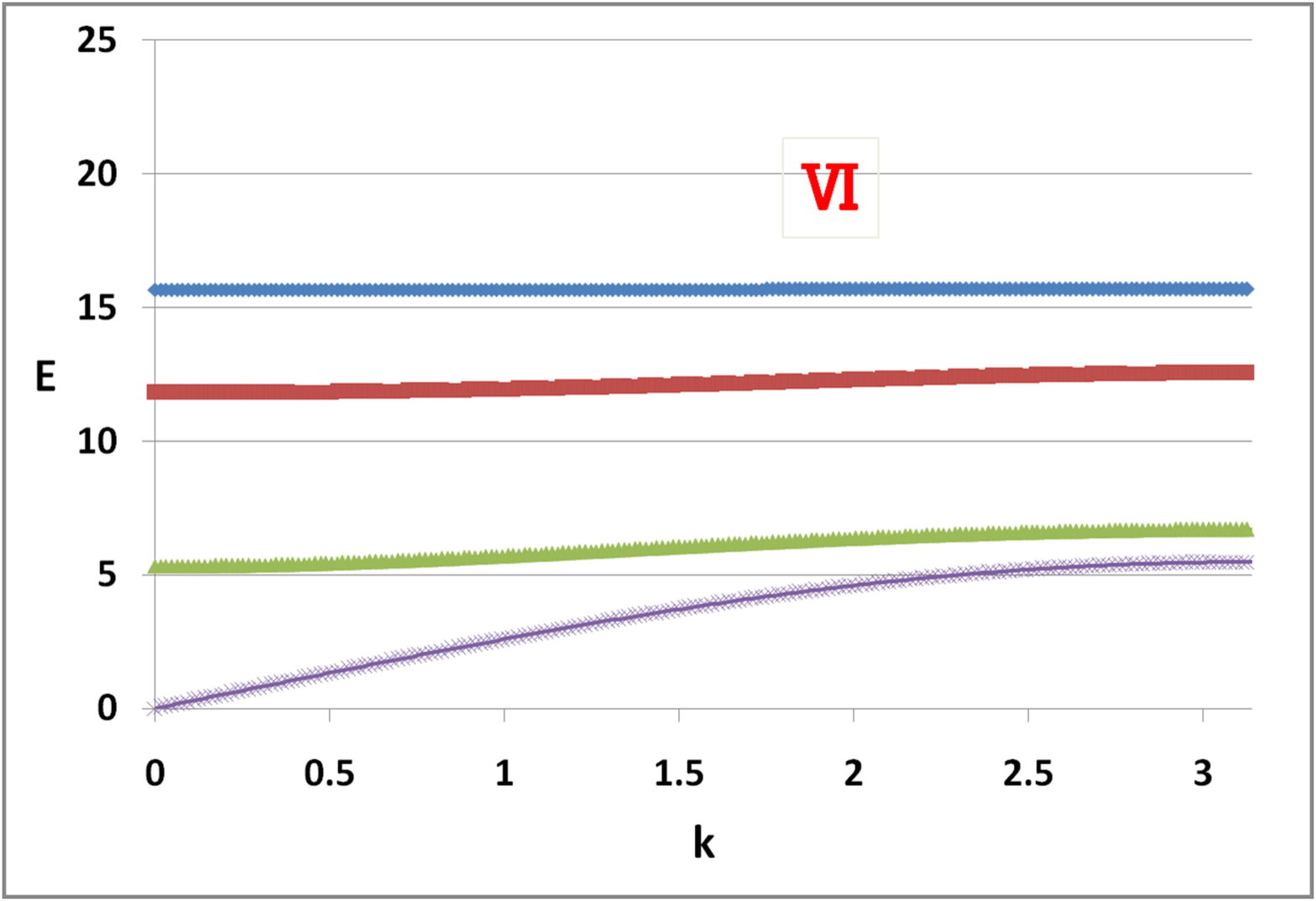}
\hspace{1.6cm} (c)
\end{minipage}\\
\begin{minipage}{0.9\hsize}
\centering
\includegraphics[width=\hsize]{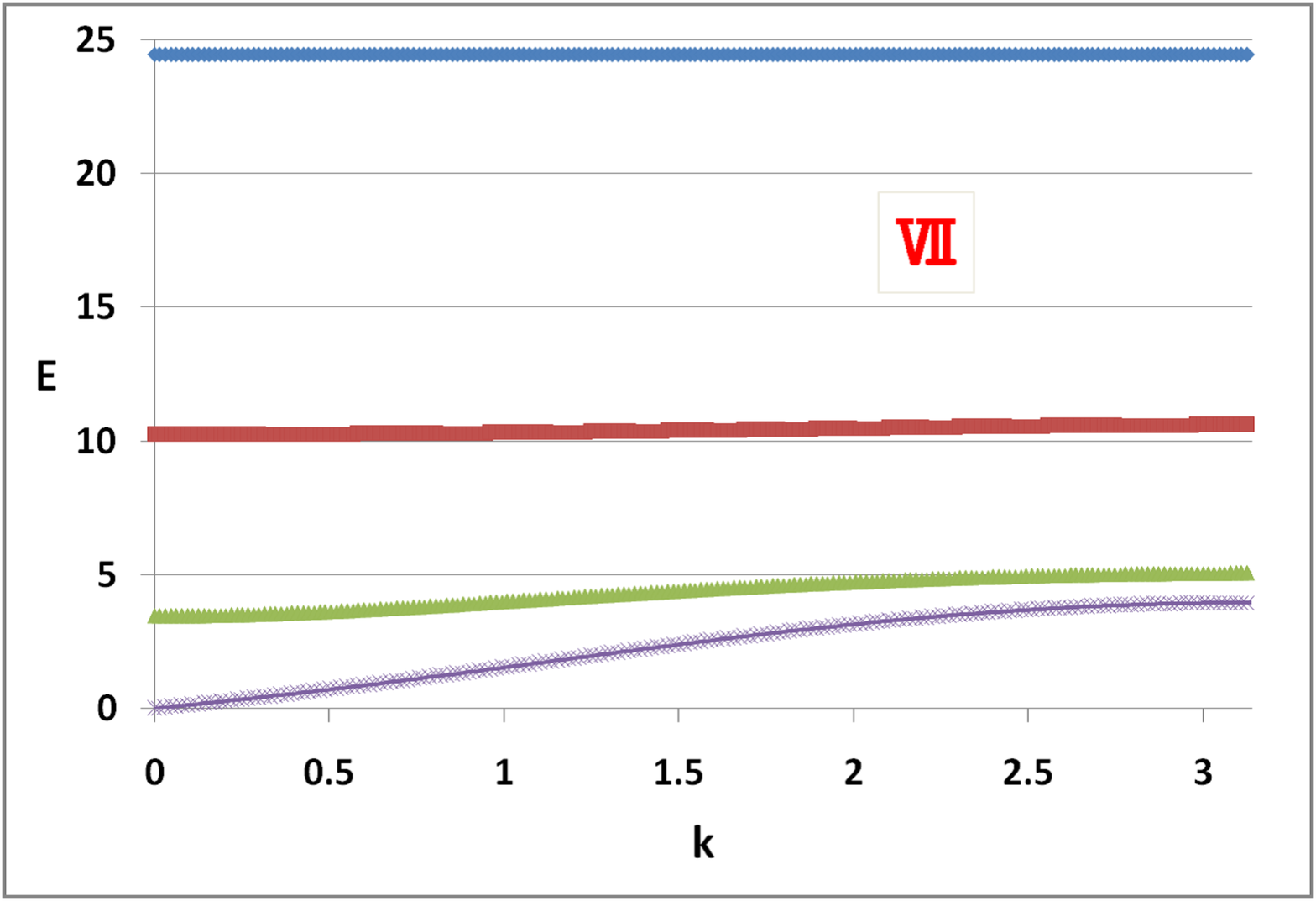}
\hspace{1.6cm} (d)
\end{minipage}
\end{tabular}
\caption{(Color online) Excitation spectra of the SF phase of the bosons in (a) region IV for $\mu_{\rm b}=1$, $\mu_{\rm f}=0.5$,
 (b) region V for $\mu_{\rm b}=6$, $\mu_{\rm f}=14$, (c) region VI for $\mu_{\rm b}=10$, $\mu_{\rm f}=6$,
and (d) region VII for $\mu_{\rm b}=10$, $\mu_{\rm f}=16$.
There is one gapless dispersive mode and three gap-containing  dispersive modes.
}
\end{center}
\end{figure}

One gapless dispersive mode and three gap-containing dispersive modes are observed. 
The gapless dispersive mode is a phase-fluctuation mode called Bogoliubov mode, while the gap-containing dispersive modes are called amplitude modes.
Regions V and VII are the bosonic SF phase with the insulator fermions; therefore, the amplitude modes shift due to interaction between the fermions and the bosons.  
\subsection{Metal phase}
Figure 4 shows the excitation spectra of the metal phase of the fermions for $\mu_{\rm b}=0.4$ and $\mu_{\rm f}=0.7$.
One gapless dispersive mode is evident. In the figure, we omitted upper level modes. 
This result is consistent with the cosine band of free fermions.
\begin{figure}[H]
\begin{center}
\centering
\includegraphics[width=\linewidth]{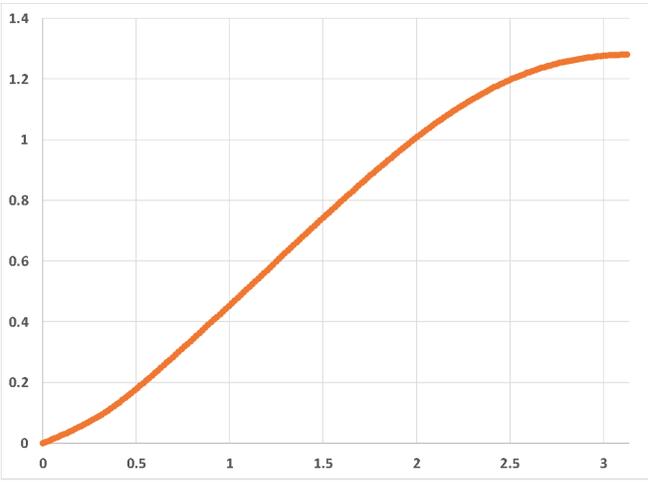}
\hspace{1.6cm} 
\caption{(Color online) Excitation spectra of the metal phase of the fermions in region VIII for
 $\mu_{\rm b}=0.4$ and $\mu_{\rm f}=0.7$. There is one gapless dispersive mode.
 }
\end{center}
\end{figure} 

\subsection{Coexisting phase}
Figure 5 shows the excitation spectra for the coexisting phase of SF bosons and metal fermions 
for (a) $\mu_{\rm b}=6$, $\mu_{\rm f}=7$, (b) $\mu_{\rm b}=6.5$, $\mu_{\rm f}=7$, 
(c) $\mu_{\rm b}=6$, $\mu_{\rm f}=7.5$, and (d) $\mu_{\rm b}=6.5$, $\mu_{\rm f}=7.5$.
The particle number densities are (a) $n_{\rm b}=0.31$, $n_{\rm f}=0.88$, (b) $n_{\rm b}=0.19$, $n_{\rm f}=0.85$, (c) $n_{\rm b}=0.32$, $n_{\rm f}=0.9$, and (d) $n_{\rm b}=0,22$, $n_{\rm f}=0.88$. 
Only the two lowest gapless dispersive modes are shown, where the lower line corresponds to the Bogoliubov mode. 
As $\mu_{\rm b}$ increases, $n_{\rm b}$ decreases and the two modes shift down.
On the other hand, as $\mu_{\rm f}$ increases, $n_{\rm f}$ increases and the two modes shift up.
In the coexisting phase, the ground state energy is determined by the Bose-Fermi and Bose-Bose interactions and the chemical potential. 
The increase of the chemical potential has an effect to increase $n_{\rm b}$ whereas low $n_{\rm b}$
is preferable to decrease the Bose-Bose interaction energy. As a result of the competition of these two, $n_{\rm b}$ decreases with the increase of  $\mu_{\rm b}$.
We also confirm these properties in the case of $U_{\rm bf}=0, 3,$ and $5$.

\begin{figure}[H]
\begin{center}
\begin{tabular}{cc}
\begin{minipage}{0.9\hsize}
\centering
\includegraphics[width=1.0\hsize]{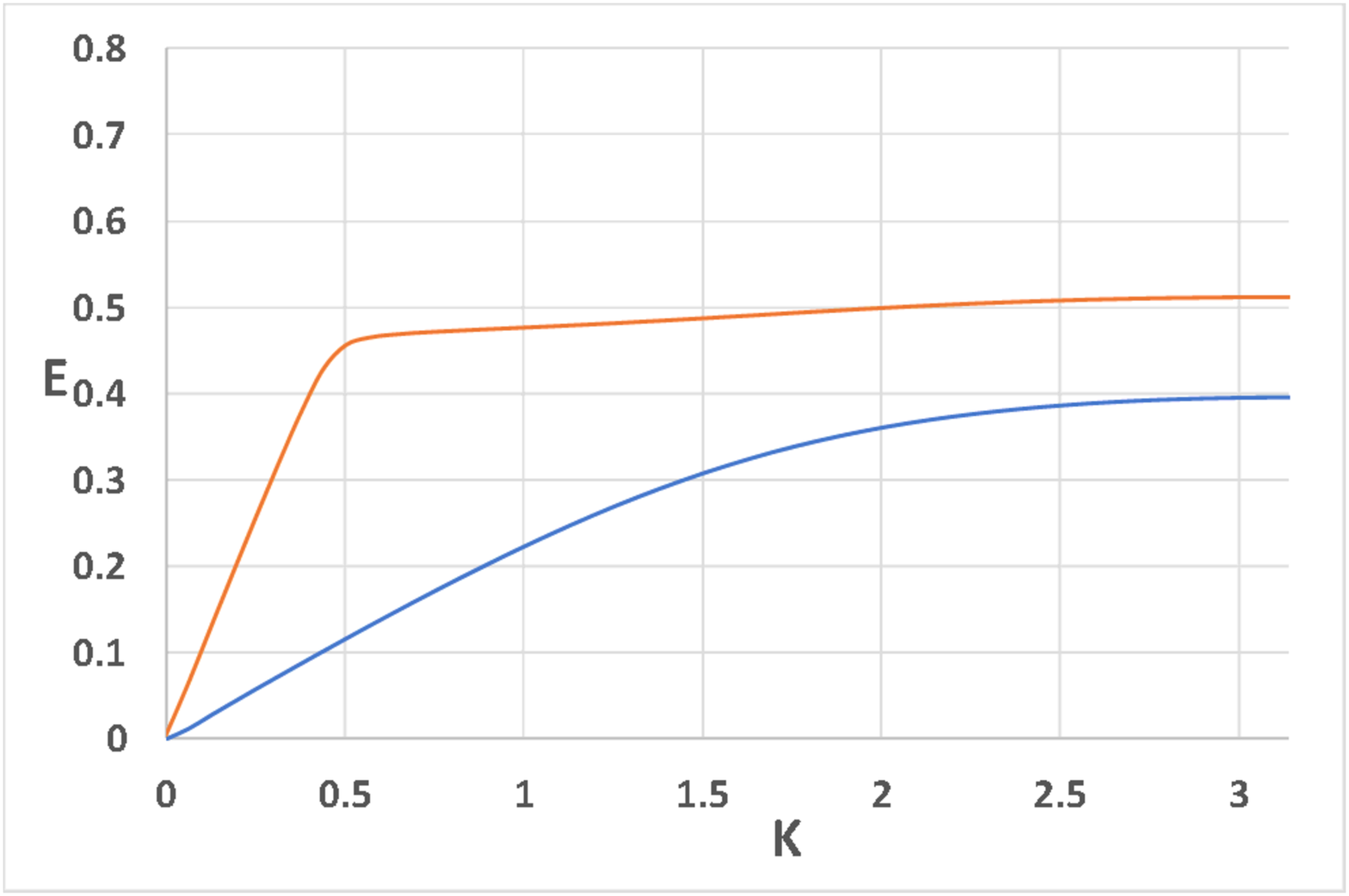}
\hspace{1.6cm} (a)
\end{minipage}\\
\begin{minipage}{0.9\hsize}
\centering
\includegraphics[width=1.0\hsize]{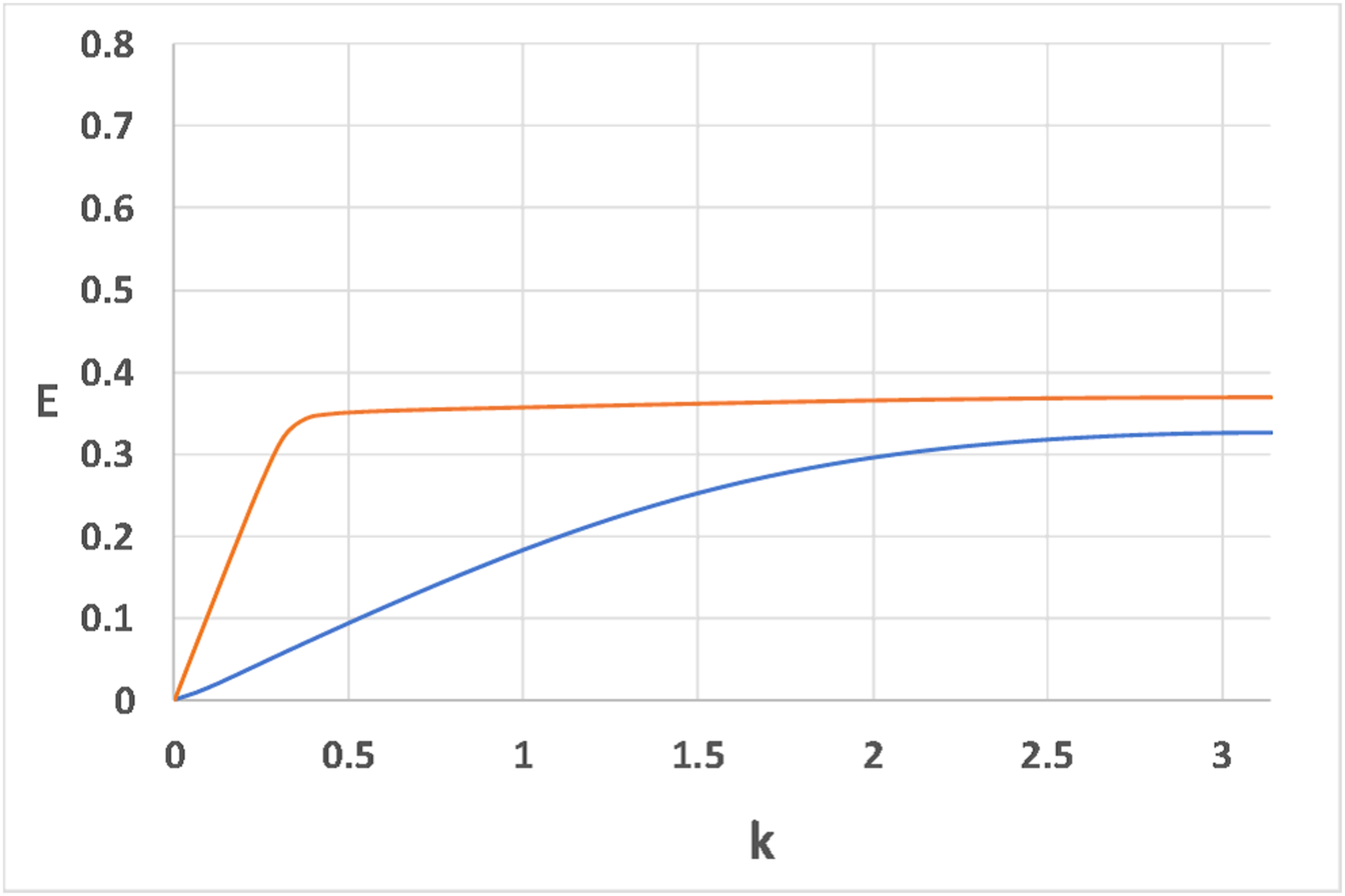}
\hspace{1.6cm} (b)
\end{minipage}\\
\begin{minipage}{0.9\hsize}
\centering
\includegraphics[width=1.0\hsize]{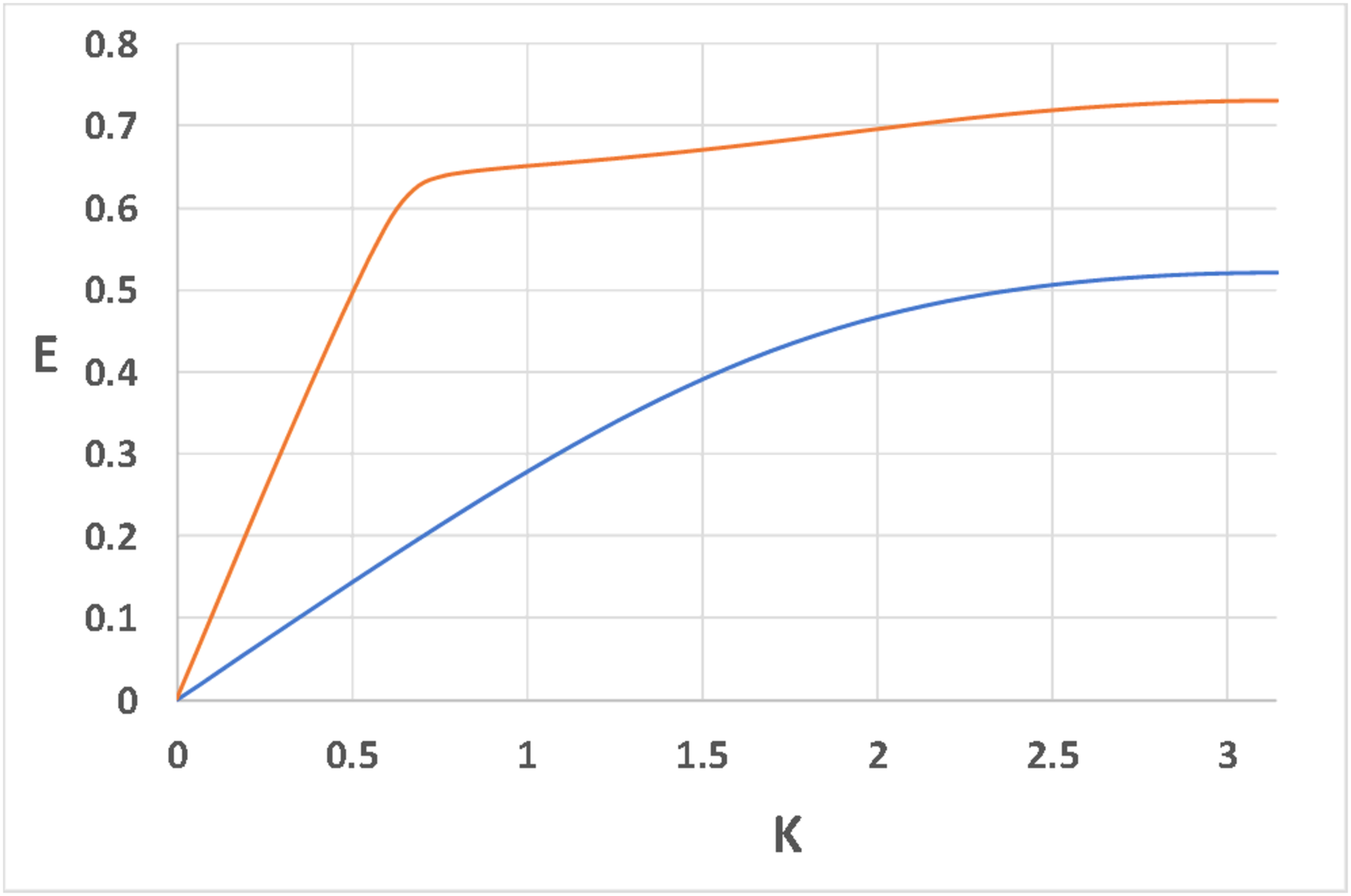}
\hspace{1.6cm} (c)
\end{minipage}\\
\begin{minipage}{0.9\hsize}
\centering
\includegraphics[width=1.0\hsize]{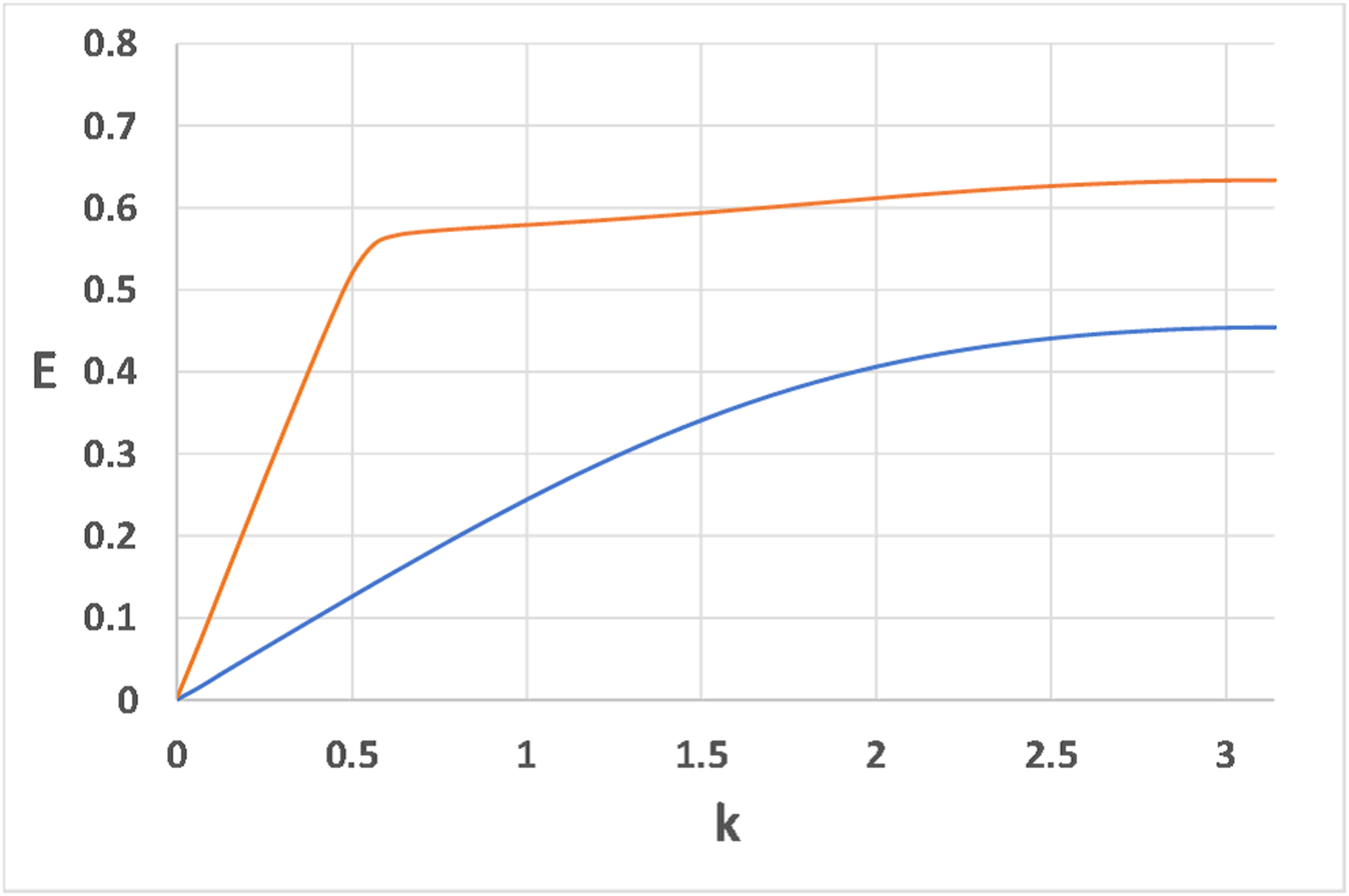}
\hspace{1.6cm} (d)
\end{minipage}
\end{tabular}
\caption{(Color online) Excitation spectra of region IX for (a) $\mu_{\rm b}=6$, $\mu_{\rm f}=7$, 
(b) $\mu_{\rm b}=6.5$, $\mu_{\rm f}=7$, (c) $\mu_{\rm b}=6$, $\mu_{\rm f}=7.5$, and (d) $\mu_{\rm b}=6.5$,
 $\mu_{\rm f}=7.5$.
The particle number densities are (a) $n_{\rm b}=0.31$, $n_{\rm f}=0.88$, (b) $n_{\rm b}=0.19$, $n_{\rm f}=0.85$, (c) $n_{\rm b}=0.32$, $n_{\rm f}=0.9$, and (d) $n_{\rm b}=0,22$, $n_{\rm f}=0.88$.
There are two gapless dispersive modes.
}
\end{center}
\end{figure} 

\section{CONCLUSION}
The ground state phase diagrams and excitations of Bose-Fermi mixtures in a 3D optical lattice were investigated using the Gutzwiller approximation.
The ground state phase diagrams were obtained that were spanned by the chemical potentials  at $\mu_{\rm b}$ and $\mu_{\rm f}$ over a wide range of $U_{\rm bb}$ and $U_{\rm bf}$. A coexisting phase of SF bosons and metal fermions was discovered. We showed the coexisting phase shrinks as the Bose-Fermi interaction $U_{\rm bf}$ increases. 
Excitation spectra were also calculated for each of the phases at $U_{\rm bb}=6$ and $U_{\rm bf}=6$ by solving the linearized equations of motion. The MI phase has two gap-containing dispersive modes, which respectively correspond to Bose particle excitation and Bose hole excitation. For the SF phase, one gapless mode called  Bogoliubov mode and three gap-containing modes called amplitude modes were obtained. The amplitude modes shifted due to interactions between the fermions and the bosons. 
In the coexisting phase of SF bosons and metal fermions, two gapless modes were obtained. These two gapless modes were also found to shift due to interactions between the bosons and the fermions.
We expect that the results presented here will stimulate further experimentations on Bose-Fermi mixtures.

\section{Acknowledgment}
\begin{acknowledgment}
This work was supported by a Kakenhi Grant-in-Aid (No. 17K14364) from the Japan Society for the Promotion of Science (JSPS).
\end{acknowledgment}

\end{document}